\documentclass[conference]{IEEEtran}
\IEEEoverridecommandlockouts
\usepackage{cite}
\usepackage{amsmath,amssymb,amsfonts}

\usepackage{graphicx}
\usepackage{textcomp}
\usepackage{xcolor}
\usepackage{balance}
\usepackage{algpseudocode}
\usepackage{algorithm}
\usepackage{amsmath}
\usepackage{mathtools}

\ifCLASSOPTIONcompsoc
\usepackage[caption=false,font=footnotesize,labelfont=sf,textfont=sf]{subfig} 
\else
\usepackage[caption=false,font=footnotesize]{subfig}
\fi

\def\BibTeX{{\rm B\kern-.05em{\sc i\kern-.025em b}\kern-.08em
		T\kern-.1667em\lower.7ex\hbox{E}\kern-.125emX}}
\begin{document}
	
	\title{Reinforcement Learning based Multi-connectivity Resource Allocation in Factory Automation Systems\\ \vspace{-1ex}} \vspace{-2ex}
	\vspace{-6ex}
	\author{\normalsize
	Mohammad Farzanullah\IEEEauthorrefmark{1}, Hung V. Vu\IEEEauthorrefmark{2} and Tho Le-Ngoc\IEEEauthorrefmark{1}\\
	\IEEEauthorblockA{\IEEEauthorrefmark{1}
		Department of Electrical \& Computer Engineering, McGill University, Montr\'{e}al, QC, Canada} 
	\IEEEauthorblockA{\IEEEauthorrefmark{2}
		Huawei Technologies Canada, Ottawa, ON, Canada} \vspace{-7ex}
		
	} 
	
	\maketitle
	
	\begin{abstract}
	
		We propose joint user association, channel assignment and power allocation for mobile robot Ultra-Reliable and Low Latency Communications (URLLC) based on multi-connectivity and reinforcement learning. 
		The mobile robots require control messages from the central guidance system at regular intervals. We use a two-phase communication scheme where robots can form multiple clusters. The robots in a cluster are close to each other and can have reliable Device-to-Device (D2D) communications. In Phase I, the APs transmit the combined payload of a cluster to the cluster leader within a latency constraint. The cluster leader broadcasts this message to its members in Phase II. We develop a distributed Multi-Agent Reinforcement Learning (MARL) algorithm for joint user association and resource allocation (RA) for Phase I. The cluster leaders use their local Channel State Information (CSI) to decide the APs for connection along with the sub-band and power level. The cluster leaders utilize multi-connectivity to connect to multiple APs to increase their reliability. The objective is to maximize the successful payload delivery probability for all robots. Illustrative simulation results indicate that the proposed scheme can approach the performance of the centralized algorithm and offer a substantial gain in reliability as compared to single-connectivity (when cluster leaders are able to connect to 1 AP). 
	\end{abstract}
	
	\vspace{0ex}
	
	
	\vspace{-1ex}
	
	\section{Introduction}
	\vspace{-1ex}
	5G is envisioned to support a variety of key services which includes Enhanced Mobile Broadband (eMBB), massive Machine-Type Communications (mMTC), and Ultra-Reliable and Low Latency Communications (URLLC). URLLC will be relevant to many emerging domains such as wireless control and automation in factory environments, Cellular Vehicle-to-Everything (C-V2X) communications, and tactile internet \cite{li20185g}. Resource allocation (RA) design for URLLC is challenging due to the high-reliability requirement (99.999\%) coupled with low latency (1 ms) requirements \cite{2018urllc}.
	
	Industrial automation is one of the key sectors where 5G aims to provide connectivity. In the industrial setting, wired connections are currently used.
	However, the factory of the future envisions to have a dynamic environment with multiple mobile robots due to which the wired solutions will not be viable. Motivated by such limitations, communications in the industrial setting is envisioned to migrate from wired to wireless communications. To guarantee a safe and efficient operation, emerging wireless technologies must support URLLC services. 
	The factory of the future will consist of multiple mobile robots requiring URLLC \cite{TR_R16}. A mobile robot is programmable and is able to perform various activities in an industry e.g., an Automated Guided Vehicle (AGV) is an autonomous vehicle to move materials within a floor. The mobile robots would require control messages from a central guidance system to avoid collisions and manage jobs. Since the robots require URLLC, an intelligent RA design is necessary.
	
	Achieving URLLC in industrial automation systems is challenging due to fading. To address this challenge, diversity has been proposed as an attractive solution to combat the fading effect in a wireless environment. The probability that multiple paths experience a deep fade simultaneously is lower, decreasing the probability of error and increasing the reliability of the system.
	In the existing literature, recent works \cite{2018urllc, popovski2019wireless} have demonstrated that multi-connectivity is a promising solution to enable URLLC. Multi-connectivity uses space and/or frequency diversity by allowing the receiver to connect to multiple transmitters (e.g., base-stations (BSs) or access-points (APs)) on either the same or different frequencies. 
	Further, extremely high data rates (correspondingly low latency) can be obtained via exploiting multi-connectivity in URLLC networks \cite{mishra2021multi}.
	\cite{2020stable} develops RA strategy in multi-user multi-cell systems with shared spectrum resources. Based on a matching theory approach, the proposed RA algorithm jointly performs link selection and sub-band scheduling in which multiple resources (sub-bands) are mapped to one user. In the context of factory automation, the multi-connectivity problem is considered, together with dynamic channel allocation for interference management in a so-called campus network where the wireless communications in factory is impaired by the external interference \cite{2020campus}. Reinforcement learning (RL) is adopted to resolve the joint user association and channel assignment problem. The authors in \cite{burgueno2021distributed} use RL in a factory setting for time-slot selection for packet transmission. However, in \cite{2020stable, 2020campus, burgueno2021distributed}, the authors do not take into consideration power allocation, which is certainly beneficial to manage the interference.
	
	Additionally, there is a strong line of work \cite{trad1_d2d,khosravirad2020} that utilizes two-hop transmission for message relaying. This stems from an observation that, due to the fading, not all robots can establish a reliable direct connection to an AP. To overcome this drawback, these works propose collaborative schemes that allow to extend the network coverage. In a factory setting, there are many devices, making it difficult to achieve URLLC with conventional broadcasting techniques. Inspired by these works, we develop a two-phase transmission strategy for a mobile robot scenario in a factory automation system. Under this strategy, the robots form multiple clusters, each consisting of one leader and multiple members. In the mobile robot scenario, the robots work in close proximity and can form reliable D2D connections. 
	The D2D connections will have stronger channels as compared to AP-to-robot connections. 
	The leader would receive control messages for its cluster from the APs in Phase I and will relay it to its members 
	in Phase II. 
	
	Moreover, the RA techniques using a central controller are not practical in a mobile URLLC setting due to stringent latency and reliability. The mobility of the robots causes fast channel variations along with the increase in latency to transmit CSI to the central controller. 
	Furthermore, traditional optimization algorithms have been used for RA in factory automation systems \cite{trad1_d2d}. However, these methods require a large number of iterations to reach a satisfying solution and are sensitive to change in system parameters.
	Recent works have adopted distributed RL approach for RA in multiple domains, e.g., in cellular communications \cite{nasir2020deep}, and C-V2X \cite{liang2019spectrum}.
	
	Reinforcement Learning (RL) is a sub-field of Machine Learning where an agent is able to learn in an interactive environment by trial and error. The intelligent agents takes an action in an environment and receives a reward.
	In RL, the aim of the agent is to maximize the cumulative reward $G_t$ it receives in the long run.
	Q-learning is a model-free RL algorithm that learns the value of an action $a$ in a state $s$. 
	It iteratively updates the action-value function for each state-action pair $(s,a)$ until they converge to the optimal action-value function $Q_{*}(s,a)$.
	The Deep Q-Learning \cite{qlearning} uses a Deep Neural Network (DNN) as a function approximator to solve for the optimal policy. The parameters of the DNN learn to approximate the Q-function in a Deep Q-Network (DQN). 

	This paper develops a two-phase scheme for control messages delivery from the APs to the robots in a factory automation system. The robots form multiple clusters,  each consisting of one leader and multiple members. During Phase I, the APs combine the messages for the whole cluster and deliver them to the cluster leader. While in Phase II, the leader broadcasts this message to its members. We develop a MARL based distributed algorithm for Phase I, where the cluster leader acts as an agent and performs joint user association, channel assignment, and power allocation. Furthermore, we utilize dual-connectivity where each leader is able to connect to a maximum of two APs at the same time in order to enhance reliability. The agents only use their local CSI and learn to optimally decide the APs to connect to along with the sub-band and power level. Our optimization objective is reliability, which is formally defined as the average probability of successful payload delivery within a required latency. Simulation results indicate that the proposed scheme outperforms the MARL single-connectivity \cite{farzanullah2022deep} and random allocation, and achieves result close to that of centralized algorithm, even outperforming it in some cases.
	
	\section{System Model and Problem Formulation}
	
	\vspace{-1ex}
	\begin{figure} [htb!]
	\centering
	\includegraphics[scale=0.27]{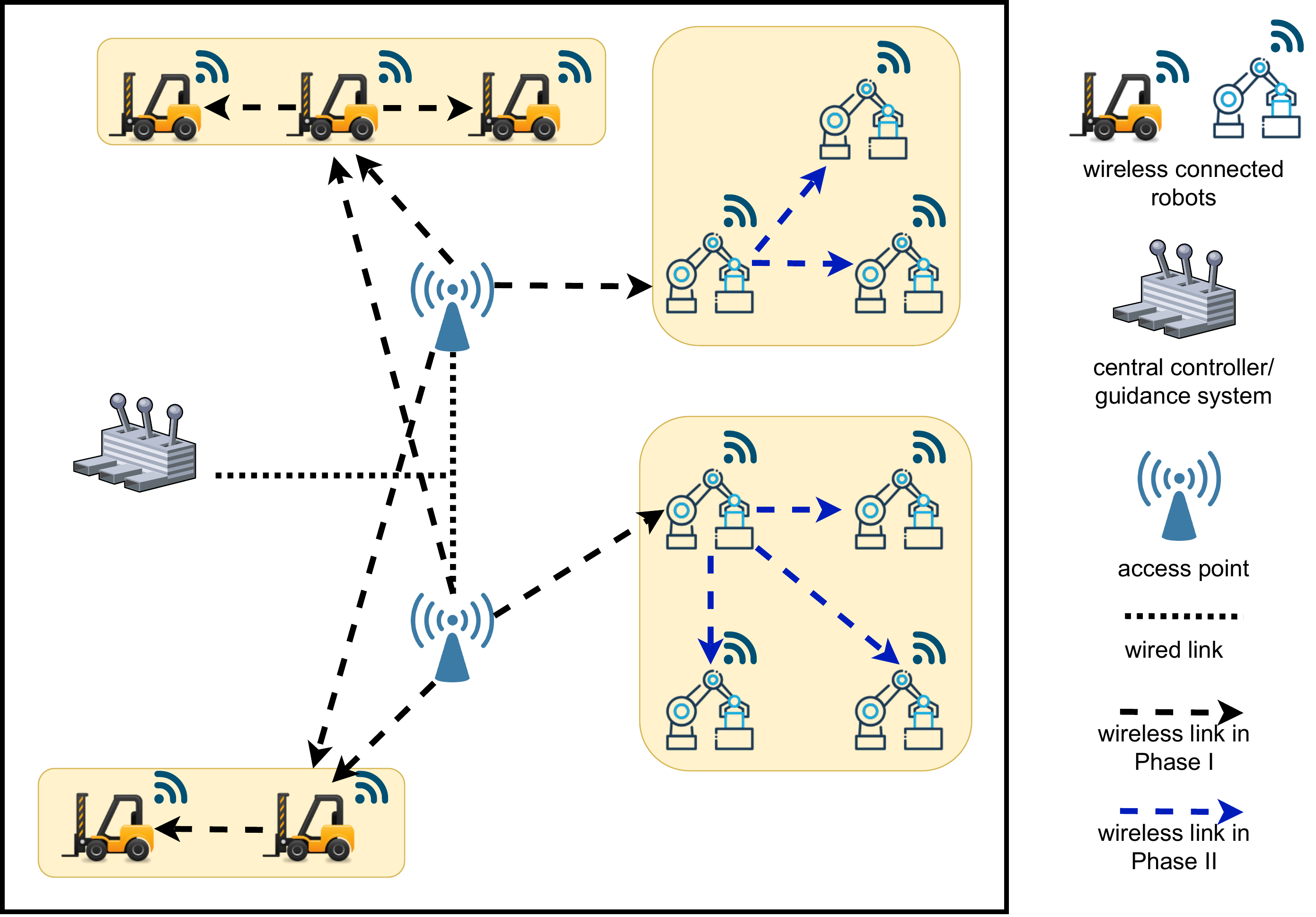}
	\vspace{-1.5ex}
	\caption{Illustrative factory/warehouse automation system with multiple clusters of mobile robots. 
	} 
	\label{Fig:SystemModel} \vspace{-4ex}
	\end{figure} 
	
	Fig.\ref{Fig:SystemModel} shows the system model, where we consider the downlink communications in a single-floor factory setting. The floor consists of multiple robots, each of which needs to a receive control message from the central controller. There are $K$ Access Points (APs) in the factory that are required to deliver control messages to all the robots. Further, we assume that there are $M$ sub-bands to communicate on. We assume that the robots are in close proximity due to which they form $N$ clusters for reliable communication. A cluster $n$ consists of one leader and $O$ members. We denote the set of members in cluster $n$ as $\mathcal{V}_n, \, n=1,\dots,N$, i.e., there are $N$ clusters with $O$ members in each cluster. Moreover, each robot can be part of single cluster, i.e., $\mathcal{V}_n  \cap  \mathcal{V}_i = \emptyset$ for $n \neq i$. We assume that the members can have a maximum distance of $d$ meters from their cluster leader. Each robot requires a payload of $B_o$ bytes within a latency of $T$ seconds. We develop a two-phase communication approach where the APs deliver the control messages to the cluster leader in the first phase and the cluster leader relays these messages to the cluster members in the second phase via point-to-multipoint D2D connections. In Phase I, the payload for the cluster is combined and the APs transmit this payload to the cluster leader. Since the payload is for the whole cluster, it consists of $B = B_o \times (O+1)$ bytes. In the second phase, the cluster leaders broadcast the received messages to its members. The set of APs, clusters and sub-bands available is denoted by $\mathcal{K}$, $\mathcal{N}$ and $\mathcal{M}$ respectively. We assume that both the APs and the robots are equipped with a single antenna. Furthermore, we assume $M = N/2$ which necessitates spectrum sharing.
	
	In Phase I, the communication is from the APs to the cluster leader. We consider multi-connectivity in Phase I, i.e., the cluster leader is able to connect to multiple APs simultaneously. The APs are connected to the central controller, which can coordinate the synchronous transmission of control messages to the cluster leader. The cluster leader needs to decide the APs it needs to connect to along with the sub-band and power level. In our work, we assume that in the case of multi-connectivity, the cluster leader can connect to a maximum of two APs, and the assigned radio-bands are located on the same sub-bands. Suppose that a cluster leader $n$ is served by the AP subset $\mathcal{K}_{nm} \subseteq \mathcal{K}$ in sub-band $m$. Then the received signal-to-interference-plus-noise ratio (SINR) at leader $n$ in sub-band $m$ is given by:
	
	\vspace{-2ex}
	\begin{align} \label{PhaseI:SINR} \small
	\text{SINR}_{n}^{(\text{c})}[m] = \frac{\sum_{k \in \mathcal{K}_{nm}} {P^{(\text{c})}_{kn} L^{(\text{c})}_{kn} g^{(\text{c})}_{kn}[m]}}{\sum_{i \in K} \sum_{j \neq n}  \rho^{(\text{c})}_{ij}[m] P^{(\text{c})}_{ij} L^{(\text{c})}_{ij} g^{(\text{c})}_{ij}[m] +\sigma^2}
	\end{align} 
	where $P_{xy}^{(\text{c})}$ refers to the power from transmitter $x$ to cluster leader $y$. $\sigma$ refers to the noise power at the receiver. Moreover, $g_{ab}^{(\text{c})}$ refers to the small-scale fading in the AP-to-leader link $ab$ and $L_{ab}^{(\text{c})}$ refers to the large-scale fading in the link $ab$. We consider that the small-scale fading power follows the Rayleigh distribution. The large-scale fading consists of both the path loss and shadowing. $\rho^{(\text{c})}_{ij}[m]$ indicates the indicator function which is set to 1 if link $ij$ reuses the sub-band $m$, and 0 otherwise.
	
	The achievable throughput of leader $n$ in sub-band $m$ is given by:
	
	\vspace{-3ex}
	\begin{align} \label{PhaseI:Rate} \small 
		R^{(\text{c})}_n = W \log_2 (1 + \text{SINR}^{(\text{c})}_{n}[m])
	\end{align}
	where $W$ is the bandwidth of each spectrum sub-band.  
	
	Phase II consists of the communication from the cluster leader to members. The cluster leader broadcasts the payload received to all of the members. We assume that each member can decode its message from the combined payload. A member $o$ will be affected by the interference from the leaders that transmit using the same sub-band. The SINR at the receiver of member $o \in O$ is given by:
	\begin{align} \label{PhaseII:SINR} \small
		\text{SINR}_{no}^{(\text{d})} = \frac{P^{(\text{d})}_{no} L^{(\text{d})}_{no} g^{(\text{d})}_{no}[m]}{ \sum_{i \neq n}  \rho^{(\text{d})}_{io}[m] P^{(\text{d})}_{io} L^{(\text{d})}_{io} g^{(\text{d})}_{io}[m] +\sigma^2}
	\end{align}
	where $P^{(\text{d})}_{no}$ ($P^{(\text{d})}_{io}$) denotes the power from transmitter $n$ ($i$) to receiver $o$ in sub-band $m$.
	
	The throughput of member $o$ of cluster $n$ is given by:
	\begin{align} \label{PhaseI:Rate} \small
		R^{(\text{d})}_o = W \log_2 (1 + \text{SINR}^{(\text{d})}_{no})
	\end{align}
	
	\subsection{Problem Formulation}
	In this work, the objective is to reliably transmit the control messages from the APs to the robots. Each robot is required to receive a payload of size $B$ bytes within a latency of $T$ seconds. Phase I consists of $T_1$ seconds during which all the cluster leaders need to receive the control messages. Mathematically, it can be stated as 	
	$
	\mathbb{P}\left(\sum_{t=t_1}^{T_1} \Delta_T R_{n}^{(\text{c})} (t) \ge B \right). 
	$
	where $\Delta_T$ is the time duration of one time slot $t$. Hence, this is an optimization problem where for each cluster leader $n$, we need to find the APs to connect to along with the sub-band and power level.
	
	For Phase II, the formulation can be mathematically stated as 	$ 
	\mathbb{P}\left(\sum_{t=t_2}^{T_2} \Delta_T R_{o}^{(\text{d})} (t) \ge B  \right) 
	$. In this Phase, we assume that the leaders transmit to the members on different resource blocks and this will be discussed in Section III. The members are close to the leaders due to which reliable transmission can be achieved. 
	
	\section{Resource Allocation using Deep Reinforcement Learning}

	\subsection{Multi-Agent Resource Allocation Algorithm}
	
	In this section, we formulate the multi-agent RL problem. The objective is to maximize the reliable transmission of control messages from the AP to the robots. Each of the cluster leader acts as an agent. At each time-step, an agent observes the environment, takes an action, and receives a reward. The agent learns to optimally allocate the resources by this trial and error method. We use RL for Phase I only.
	
	The multi-agent RL problem consists of two phases that are training and testing. The training is centralized and a common reward is used to ensure collaboration between the agents. This common reward is available to all the agents during the training phase. Based on the reward, the agents train their distinct DQN to fulfill the common objective. If each of the robots receives a separate reward, it would become a competitive game where each of them will try to maximize their own throughput, resulting in increased interference. Meanwhile, during the testing phase, the agents use the trained DQN to select the action. 
	The state space, action space, and the reward design are outlined below:
	\subsubsection{State Space}

	The state space of the agent consists of the measurements from the last time-step. Let $\mathcal{Z}_n (t)$ denote the state of agent $n$ at time-step $t$. The state space consists of the following measurements: the direct channels from all the APs $k \in K$ to the cluster leader $n$, i.e., $\{L^{(\text{c})}_{kn}, g^{(\text{c})}_{kn}\}_{k \in K}$, the aggregate interference and noise powers at the agent $n$, i.e., $\{{\sum_{i \in K} \sum_{j \neq n}  \rho^{(\text{c})}_{ij}[m] P^{(\text{c})}_{ij} L^{(\text{c})}_{ij} g^{(\text{c})}_{ij}[m] +\sigma^2}\}$, the remaining payload and time limitation after the current time-step, the training iteration number $e$, and the probability of random action selection $\epsilon$ of agent $n$. 

	\subsubsection{Action Space}
	
	Each agent $n$ has an identical action space. The action space consists of the APs selected along with the sub-band and power level. We limit the connection of the cluster leader to a maximum of 2 APs at time-step $t$. 
	Moreover, to reduce the complexity, we assumed that in case the cluster leader decides to connect to 2 APs, it would be assigned the same sub-band. 
	This reduces the complexity of the DQN, though at the costs of higher interference.
	The power level is divided into discrete levels in the range $[0, P_d]$ where $P_d$ is the maximum allocated power.
	
	\subsubsection{Reward Design}
	
	A common reward is used between the agents to ensure collaboration between them. We design the reward function correlating to the final objective, i.e., maximizing the successful payload delivery probability to the robots. The reward is designed as the sum of the rate of all of the cluster leaders. The rate at a cluster leader $n$ at time-step $t$ is thus defined as:
	
	\vspace{-2ex}
	\begin{align} \label{reward1} \small
	U_{n}^{(\text{c})}(t) = \left\{ \begin{array}{l}
		R_{n}^{(\text{c})} (t), \quad \text{if} \, B_{n}(t) \ge 0, \\
		U, \,\,\,\quad\quad\quad\,\,  \text{otherwise}. 
	\end{array} \right.
	\end{align}

	where $U$ needs to be empirically adjusted. Experiments show that the value of $U$ should be greater than the achievable rate at leader $n$, and lesser than two times of achievable rate. Meanwhile, $B_n$ is the remaining payload for cluster leader $n$ at time-step $t$. The common reward at time-step $t$ is defined as:
	
	\vspace{-5ex}
	\begin{align} \label{reward2} \small
		r_t = \sum_{n=1}^{N}  U_{n}^{(\text{c})} (t)
	\end{align}

	\subsection{Training Algorithm and Testing Strategy}

	There is a time limitation of $T = T_1 + T_2$ seconds to complete the transmission to all of the robots. The DQN is designed as an episodic setting, where each episode is of time duration $T$. The position of the robots and the large-scale fading is updated at the start of each episode, while the small-scale fading is updated at each time-step. At each time-step, the variation in small-scale fading changes the dynamics of the system, causing the agents to adjust their actions. The training phase consists of only Phase I. Deep Q-Learning is used to train the agents and the algorithm is outlined in Algorithm 1.
	
	\vspace{-1ex}
	\begin{algorithm}   \small
	\caption{\small Training Algorithm}	
	\begin{algorithmic}[1] 
		\State Initiate the environment
		\State Initiate the parameters $\theta$ for DNN of all cluster leaders 
		\For{each episode}
		\State Update the device locations and large-scale fading
		\For{each time-step $t$}
		\State{Update small-scale fading}
		\For{each cluster leader $n$}
		\State Observe the state $s_t$				
		\EndFor
		\State The leaders take action according to the $\epsilon$-greedy policy and obtain a common reward $r_t$ 
		\For{each cluster leader $n$}
		\State Observe the next state $s_{t+1}$		
		\State Store $<s_t, a_t, r_{t}, s_{t+1}>$ in the replay memory				
		\EndFor
		\EndFor
		\For{each cluster leader $n$}
		\State Sample a mini-batch $X$ from the replay memory
		\State Train the deep Q-network for cluster leader $n$ using $X$			
		\EndFor		
		\EndFor 
	\end{algorithmic}
 
	\end{algorithm} 
	
	\vspace{-2ex}
	The testing phase consists of both Phase I and Phase II. In Phase I, the agents observe the state space $\mathcal{Z}_n (t)$ and use the trained DNN to output the optimal action at each time-step. After Phase I, a fixed scheme is used for Phase II where each cluster leader communicates to its members in different time-frequency slots, i.e., on separate sub-band and time-step combination, which eliminates the need for interference management. The robots can reliably communicate using this scheme since the distance between the cluster leader and members is less.
	
	\vspace{-1ex}
	
	\section{Illustrative Results}
	\vspace{-1ex}
	We consider downlink communication for a factory automation system that consists of multiple robots. The robots need to periodically receive the control messages from the central guidance system. The robots can perform multiple operations in a factory and can have high mobility. Further, we assume robots form clusters, and each cluster works together to complete a specific task.
	
	\vspace{-0.25ex}
	We consider a factory floor setting of dimension $40\times40 \text{m}^2$. There are 4 APs in the factory floor setting with coordinates $(10,10)$, $(10,30)$, $(30,10)$ and $(30,30)$ assuming the south-east corner to be the $(0,0)$ coordinate. Further, there are $N$ clusters on the floor with $O$ members in each. 
	Further, we assume that the members are at a maximum distance of $d$ meters from the leader.
	The robots can move with a velocity of 1 m/s and each cluster moves either to the north, south, east or west with an equal probability. We consider both large-scale and small-scale fading in our simulations. The large-scale and small-scale fading is updated every 1 ms and 0.1667 ms respectively. For large-scale fading, we consider both path loss and shadowing. For path loss we consider the A1 indoor scenario of the WINNER II channel model \cite{winner}. 
	We assume log-normal shadowing with a standard deviation of 3 dB. Moreover, we assume Rayleigh fading for small-scale fading. 
	The simulation parameters are highlighted in Table \ref{Table:Para}.
	\vspace{0ex}
	\begin{table} 
		\caption{Simulation Parameters}	\vspace{-1ex}
		\centering %
		\begin{tabular}{|c | c|}				
			\hline	
			Carrier frequency & $3$ GHz \\ \hline	
			Bandwidth of each sub-band & $1$ MHz \\ \hline 	
			Network area & $40 \times 40$ $\text{m}^\text{2}$ \\ \hline 
			Number of clusters $N$ & [4, 6, 8] \\ \hline
			Number of members in each cluster $O$ & 4 \\ \hline
			Robot velocity & 1 m/s \\ \hline
			Transmit power of AP to cluster leader & [-100, 20, 25, 30] dBm \\ \hline
			Noise figure & 5 dB \\ \hline
			Noise PSD & -169 dBm/Hz \\ \hline
			Latency $T$ & 1 ms \\ \hline
			Latency of Phase I $T_1$ & 0.667 ms \\ \hline
			Latency of Phase II $T_2$ & 0.333 ms \\ \hline
			Robot payload $B_o$ & [20, 100] bytes \\ \hline
			Max distance between leader and members $d$ & 3 m \\ \hline
		\end{tabular} \vspace{-5ex}
		\label{Table:Para}
	\end{table}
	\vspace{0ex}

	We used Python to develop the simulation, and implemented the DQN using the Tensorflow framework. There are 3 hidden layers in the DNN for DQN, with 83, 41, and 20 nodes respectively. RMSProp was used as the optimizer with a learning rate of 0.001 while rectified linear unit (ReLU) was used as the activation function. The training phase consists of 6000 episodes, where each episode consists of 6 time-steps. Meanwhile, the testing phase consists of 1000 episodes. For URLLC applications, we consider a sub-frame of 1 ms which is further divided into 6 sub-slots \cite{urllc5g}. 
	We use the $\epsilon$-greedy strategy during the training phase, where the value of $\epsilon$ was linearly reduced from 1 to 0.02 for 80\% of the training and was fixed at 0.02 after that. The discount factor for DQN is set at 0.9. The value of $U$ defined in Eq. \ref{reward1} was set as 40. 
	In the training phase, we fixed the value of robot payload size $B_o$ as 100 bytes, while we varied the payload size between 20 bytes and 100 bytes for the testing phase \cite{5g_acia}.
	
	

We develop five benchmarks for comparison with our proposed strategy:	
	\begin{itemize}
		\item Random Allocation: Each cluster leader connects to the nearest AP and randomly selects the sub-band and power level.
		\item Centralized Algorithm: Each cluster leader connects to the nearest AP and exhaustively searches for the sub-band and power level that maximizes the sum-rate.
		\item Single-connectivity MARL: We run the RL algorithm with the restriction that each of the cluster leaders can connect to just one AP.
		\item Greedy single-connectivity: Each cluster leader using the single best link available to transmit at maximum power.
		\item Greedy multi-connectivity: Each cluster leader using the two best links available to transmit at maximum power.
	\end{itemize}
	
%
	
	 \vspace{-1ex}
	\begin{figure} [htb!]
	\centering
	\includegraphics[scale=0.59]{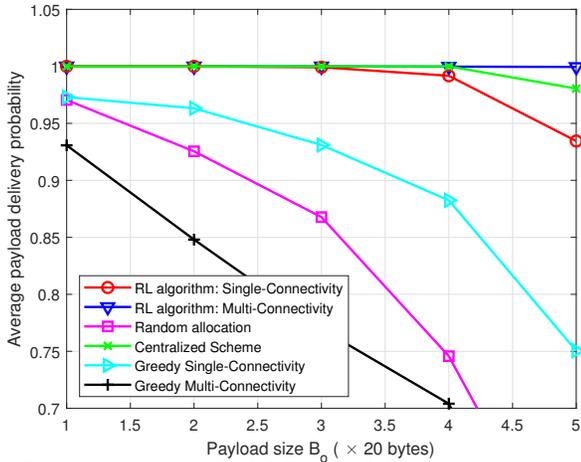}
	\vspace{-3ex}
	\caption{Average robot payload delivery probability versus payload size for $N$=4}
	\label{Fig:success4} \vspace{-2ex}
	\end{figure}

	Fig. \ref{Fig:success4} shows the average successful payload delivery probability versus the payload size. The proposed algorithm was trained for $N$ = 4 and the payload size $B_o$ was varied during the testing phase. Our proposed algorithm achieves a successful payload delivery probability of 1 for payload size up to 60 bytes, and more than 0.999 for up to 100 bytes, outperforming both the centralized algorithm and the RL algorithm with single-connectivity.
	Moreover, it is to be noted that for the greedy approach, which is a naive strategy, single-connectivity performs better than multi-connectivity. This is due to higher interference when all of the cluster leaders attempt to use multiple links to complete their transmission. However, for RL, there is a substantial improvement in performance by utilizing multi-connectivity. This shows that multi-connectivity can yield a greater benefit only with an intelligent RA design.
	
%
	
	\begin{figure} [htb!]
	\centering \vspace{-2ex}
	\includegraphics[scale=0.54]{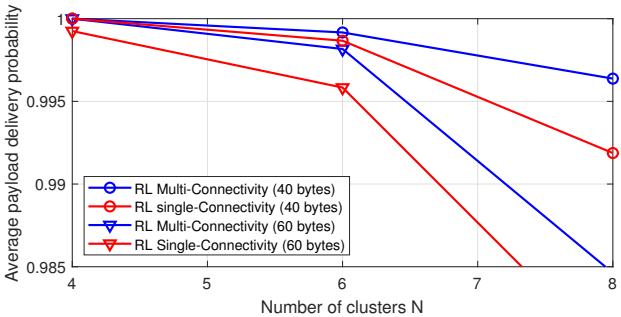}
	\vspace{-4ex}
	\caption{Average robot payload delivery probability versus number of clusters}
	\label{Fig:clusters} \vspace{-0ex}
\end{figure} 
	\vspace{-2ex}

	Fig. \ref{Fig:clusters} shows the effect of the number of clusters $N$ on the performance for the payload $B_o$ of 40 bytes and 60 bytes. For all cases, the average robot payload delivery probability is higher than 0.995, and degraded when $N>6$, mainly due to the increasing interference among the cluster leaders. Furthermore, the training for MARL becomes more challenging as we increase the number of agents. 
	
	\vspace{0ex}
	\section{Conclusion}
	\vspace{-0.5ex}
	The paper utilized multi-connectivity based user association and resource allocation for a mobile robot scenario in a factory automation system. A two-phase scheme was considered for message relaying, where in Phase I the APs deliver control messages to the cluster leader and in Phase II the cluster leader broadcasts the messages to its members. For Phase I, we developed a distributed MARL algorithm for joint user association, channel assignment, and power allocation. For user association, each cluster leader was able to connect to multiple APs to enhance transmission reliability. 
	The proposed algorithm achieved results close to that of the centralized algorithm. Moreover, it was shown that utilizing multi-connectivity offers substantial gain in reliability.
	In the future, cluster formation and cluster leader selection can be optimized, or multiple antennas at AP can be considered for beam-forming.
	
	\section*{Acknowledgment}
	This work was supported in part by the Natural Sciences and Engineering Research Council of Canada and in part by Huawei Technologies Canada.
	 \vspace{-0.5ex}
	\balance
	\bibliographystyle{IEEEtran}
	\bibliography{PaperMC_20220825}

\begin{thebibliography}{10}
\providecommand{\url}[1]{#1}
\csname url@samestyle\endcsname
\providecommand{\newblock}{\relax}
\providecommand{\bibinfo}[2]{#2}
\providecommand{\BIBentrySTDinterwordspacing}{\spaceskip=0pt\relax}
\providecommand{\BIBentryALTinterwordstretchfactor}{4}
\providecommand{\BIBentryALTinterwordspacing}{\spaceskip=\fontdimen2\font plus
\BIBentryALTinterwordstretchfactor\fontdimen3\font minus
  \fontdimen4\font\relax}
\providecommand{\BIBforeignlanguage}[2]{{%
\expandafter\ifx\csname l@#1\endcsname\relax
\typeout{** WARNING: IEEEtran.bst: No hyphenation pattern has been}%
\typeout{** loaded for the language `#1'. Using the pattern for}%
\typeout{** the default language instead.}%
\else
\language=\csname l@#1\endcsname
\fi
#2}}
\providecommand{\BIBdecl}{\relax}
\BIBdecl

\bibitem{li20185g}
Z.~Li, M.~A. Uusitalo, H.~Shariatmadari, and B.~Singh, ``{5G URLLC:} {Design}
  challenges and system concepts,'' in \emph{2018 15th International Symposium
  on Wireless Communication Systems (ISWCS)}.\hskip 1em plus 0.5em minus
  0.4em\relax IEEE, 2018, pp. 1--6.

\bibitem{2018urllc}
M.~Bennis, M.~Debbah, and H.~V. Poor, ``Ultrareliable and low-latency wireless
  communication: Tail, risk, and scale,'' \emph{Proceedings of the IEEE}, vol.
  106, no.~10, pp. 1834--1853, Oct. 2018.

\bibitem{TR_R16}
``3rd {Generation} {Partnership} {Project}; {Technical} {Specification} {Group}
  {Services} and {System} {Aspects}; {Study} on {Communication} for
  {Automation} in {Vertical} {Domains} (release 16),'' 3GPP, Tech. Rep., Jul.
  2020.

\bibitem{popovski2019wireless}
P.~Popovski, {\v{C}}.~Stefanovi{\'c}, J.~J. Nielsen, E.~De~Carvalho,
  M.~Angjelichinoski, K.~F. Trillingsgaard, and A.-S. Bana, ``Wireless {Access}
  in {Ultra-Reliable} {Low-Latency} {Communication} ({URLLC}),'' \emph{{IEEE}
  Trans. Commun.}, vol.~67, no.~8, pp. 5783--5801, Aug. 2019.

\bibitem{mishra2021multi}
P.~Mishra, S.~Kar, V.~Bollapragada, and K.-C. Wang, ``Multi-connectivity using
  nr-dc for high throughput and ultra-reliable low latency communication in 5g
  networks,'' in \emph{2021 IEEE 4th 5G World Forum (5GWF)}.\hskip 1em plus
  0.5em minus 0.4em\relax IEEE, 2021, pp. 36--40.

\bibitem{2020stable}
T.~H{\"o}{\ss}ler, P.~Schulz, E.~A. Jorswieck, M.~Simsek, and G.~P. Fettweis,
  ``Stable {Matching} for {Wireless} {URLLC} in {Multi-Cellular}, {Multi-User}
  {Systems},'' \emph{{IEEE} Trans. Commun.}, vol.~68, no.~8, pp. 5228--5241,
  Aug. 2020.

\bibitem{2020campus}
B.~Khodapanah, T.~H{\"o}{\ss}ler, B.~Yuncu, A.~N. Barreto, M.~Simsek, and
  G.~Fettweis, ``Coexistence {Management} for {URLLC} in {Campus} {Networks}
  via {Deep} {Reinforcement} {Learning},'' in \emph{IEEE Wireless Commun.
  Networking Conf. (WCNC)}, 2020, pp. 1--6.

\bibitem{burgueno2021distributed}
J.~Burgue{\~n}o, R.~Adeogun, R.~L. Bruun, C.~S.~M. Garc{\'\i}a, I.~de-la
  Bandera, and R.~Barco, ``Distributed deep reinforcement learning resource
  allocation scheme for industry 4.0 device-to-device scenarios,'' in
  \emph{2021 VTC2021-Fall}.\hskip 1em plus 0.5em minus 0.4em\relax IEEE, 2021,
  pp. 1--7.

\bibitem{trad1_d2d}
L.~Liu and W.~Yu, ``A {D2D-based} {Protocol} for {Ultra-Reliable} {Wireless}
  {Communications} for {Industrial} {Automation},'' \emph{{IEEE} Trans.
  Wireless Commun.}, vol.~17, no.~8, pp. 5045--5058, Aug. 2018.

\bibitem{khosravirad2020}
S.~R. Khosravirad, H.~Viswanathan, and W.~Yub, ``Exploiting {Diversity} for
  {Ultra-Reliable} and {Low-Latency} {Wireless} {Control},'' \emph{{IEEE}
  Trans. Wireless Commun.}, Sep. 2020.

\bibitem{nasir2020deep}
Y.~S. Nasir and D.~Guo, ``Deep {Reinforcement} {Learning} for {Joint}
  {Spectrum} and {Power} {Allocation} in {Cellular} {Networks},'' \emph{arXiv
  preprint arXiv:2012.10682}, 2020.

\bibitem{liang2019spectrum}
L.~Liang, H.~Ye, and G.~Y. Li, ``Spectrum {Sharing} in {Vehicular} {Networks}
  {Based} on {Multi-Agent} {Reinforcement} {Learning},'' \emph{{IEEE} J. Sel.
  Areas Commun.}, vol.~37, no.~10, pp. 2282--2292, Oct. 2019.

\bibitem{qlearning}
C.~J. C.~H. Watkins, ``Learning from delayed rewards,'' Ph.D. dissertation,
  King's College, Cambridge United Kingdom, May 1989.

\bibitem{farzanullah2022deep}
M.~Farzanullah, H.~V. Vu, and T.~Le-Ngoc, ``Deep reinforcement learning for
  joint user association and resource allocation in factory automation,'' in
  \emph{IEEE Wireless Commun. Networking Conf. (WCNC)}.\hskip 1em plus 0.5em
  minus 0.4em\relax IEEE, 2022, pp. 2059--2064.

\bibitem{winner}
P.~Kyosti, ``{WINNER} {II} channel models,'' \emph{IST, Tech. Rep. IST-4-027756
  WINNER II D1. 1.2 V1. 2}, 2007.

\bibitem{urllc5g}
T.~Fehrenbach, R.~Datta, B.~G{\"o}ktepe, T.~Wirth, and C.~Hellge, ``{URLLC}
  services in {5G} low latency enhancements for {LTE},'' in \emph{2018 IEEE
  88th Vehicular Technology Conference (VTC-Fall)}, 2018, pp. 1--6.

\bibitem{5g_acia}
``{5G} for {Connected} {Industries} and {Automation} ({White} {Paper-Second}
  {Edition}),'' 5G-ACIA, Feb. 2019.

\end{thebibliography}
	\balance \balance
	
	
\end{document}